\documentclass[a4paper, amsfonts, amssymb, amsmath, reprint, showkeys, twoside,superscript address]{revtex4-2}

\newcommand{\be}{\begin{equation}}
\newcommand{\ee}{\end{equation}}

\usepackage{bm,amssymb,amsmath}
\usepackage{graphicx,color,upgreek}
\usepackage{float}
\usepackage{xcolor}
\usepackage{relsize}
\usepackage{comment}
\usepackage{enumitem}
\usepackage[english]{babel}
\usepackage{letltxmacro}
\LetLtxMacro{\originaleqref}{\eqref}
\renewcommand{\eqref}{Eq.~\originaleqref}

\begin{document}

\vspace*{+1.00cm}

\title{Necessary conditions for the Markovian Mpemba effect}
 \author{Ido Avitan}
 \affiliation{Department of Physics, Technion-Israel Institute of Technology, Haifa 3200003, Israel}
 \author{Roee Factor}
 \affiliation{Department of Physics, Technion-Israel Institute of Technology, Haifa 3200003, Israel}
 \author{David Gelbwaser-Klimovsky}
 \email{\tt dgelbi@technion.ac.il}
 \affiliation{Schulich Faculty of Chemistry and Helen Diller Quantum Center, Technion-Israel Institute of Technology, Haifa 3200003, Israel}

\title{Necessary conditions for the Markovian Mpemba effect}

\begin{abstract}
The Mpemba effect is a thermodynamic anomaly in which a system farther away in temperature from equilibrium thermalizes before one that is initially closer.  The effect has been experimentally observed across a wide range of systems, including water, colloids, and trapped ions. It has recently been the focus of numerous studies aimed at understanding its  mechanisms and developing multiple applications.  Despite extensive work in the field, clearly determining which types of systems exhibit the Mpemba effect remains an open question. To address this, we derive simple necessary conditions on the 
transition rates for the Mpemba effect in a Markovian 3-level system and show that they can be applied to study the Mpemba effect in an N-level system. Multiple time scales govern thermalization in these systems. This allows the evolution to occur more quickly across larger temperature differences, explaining the Mpemba effect.  We apply our protocol to evaluate which types of systems exhibit the Mpemba effect and, in doing so, explain why the Mpemba effect in Markovian systems remains a thermodynamic anomaly. In particular, due to the maximum entropy principle, our conditions allow us to discard the sub-Ohmic and Ohmic spectra.  The latter describes a wide range of physical and chemical phenomena, which will not exhibit the Mpemba effect. Moreover, our results provide a clear path to determine the minimal physical requirements for the Mpemba effect, and we apply them to understand its underlying mechanisms better. Finally, our protocol could help identify relevant parameters for experiments, numerical simulations and diverse applications.  
\end{abstract}

\maketitle

The Mpemba effect is a thermodynamic anomaly in which a system farther from the equilibration temperature thermalizes faster than another that is initially closer to it. Its first observation dates back to Aristotle, who described situations in which hot water freezes faster than cold water \cite{webster1923aristotle}. The effect was rediscovered in the 20th century by accident by Mpemba, a secondary school student \cite{mpemba1969cool}, who questioned scientific dogma. His work initiated a still ongoing debate regarding the existence of the effect and its underlying physical mechanisms \cite{teza2026speedups,burridge2016questioning,bechhoefer2021fresh}.
	
The Mpemba effect is not exclusive to water. It has been experimentally observed in a wide range of physical systems such as colloids \cite{kumar2020exponentially}, spin glasses \cite{baity2019mpemba}, granular gases \cite{biswas2020mpemba} and trapped ions \cite{aharony2024inverse}.  Moreover, inspired by the original Mpemba effect, several works have studied ``accelerated'' or anomalous thermalization in diverse scenarios such as states with quantum coherence \cite{ivander2023hyperacceleration,moroder2024thermodynamics}, non-Markovian baths \cite{strachan2025non}, driven systems \cite{degunther2022anomalous}, systems that violate detailed balance \cite{blum2025thermalization},   exponential relaxation \cite{klich2019mpemba}, and repeated interactions setups \cite{ramon2025thermal}. The extensive work on the Mpemba and Mpemba-like effects has recently been reviewed \cite{teza2026speedups,ares2025quantum}. Furthermore, the study of the Mpemba effect has multiple and diverse applications, such as improving heating/cooling protocols \cite{gal2020precooling}, minimizing dissipation \cite{walker2023optimal}, boosting sensors' performance \cite{pagare2024mpemba}, slowing biochemical reactions \cite{hatakeyama2024enzymatic}, and improving Monte Carlo algorithms \cite{klich2018solution}.

 Determining which systems exhibit the Mpemba effect is challenging. First steps in this direction were taken in  \cite{lu2017nonequilibrium} for  Markovian systems. In that work, it was shown that the Mpemba effect is related to the overlap between the initial state and the dynamics' slow mode. The Mpemba effect requires this overlap to have 
 a non-monotonic dependence on the initial state temperature. Nevertheless, it remains unclear how to determine the physical conditions under which this non-monotonicity arises.  This is especially true for discrete systems, such as quantum systems, where the energetic landscape cannot be controlled and, when the number of states is large, analytical calculations of the dynamics are difficult to obtain. This has motivated the use of machine learning to identify systems that exhibit the Mpemba effect \cite{amorim2023predicting}. Nevertheless,  these methods have their own limitations and have been applied only to the Ising model. The lack of a general knowledge of which systems exhibit the Mpemba effect 
	hinders progress in understanding its physical mechanisms, in conducting experiments and numerical simulations, and in exploiting its potential applications. 
	 
      In this paper, we tackle the challenge of identifying which systems exhibit the Mpemba effect by studying the thermalization of quantum systems, which is governed by the Pauli rate equation.   
	We derive simple necessary conditions on the transition rates for the Mpemba effect in 3-level systems (3LSs). Furthermore, we \emph{analytically prove}  for baths at zero temperature, and numerically show for the non-zero temperature case, that the 3-level systems conditions can be used to study the Mpemba effect in N-level systems. This allows us to study the Mpemba effect in large systems using the analytical expressions for 3-level systems. We apply our protocol to identify the physical mechanisms underlying the Mpemba effect and to evaluate which types of systems exhibit it.   Among other things, we show that the standard model of a memory-less bath, whose bath spectrum is effectively flat, does not produce the Mpemba effect. This is a consequence of the maximum entropy principle which prevents the transition rates from having the  correct asymmetry required for the Mpemba effect. Moreover, our inequalities allow us to precisely characterize the asymmetry required for transition rates to produce the Mpemba effect and to discard sub-Ohmic and Ohmic spectra.  The latter is commonly used to model a wide range of physical and chemical phenomena \cite{grifoni1995cooperative,nakamura2024gate}, which, based on our results, will not exhibit the Mpemba effect.

    \section*{N-Level system thermalization} 
    
   We consider a non-degenerate $N-$level system with energy  levels $\{E_1<E_2,...<E_N \}$.  The system interacts with a thermal bath at inverse temperature $\beta_{b}$. To study the Mpemba effect in this system, we need to understand the time scales of the
 dynamics. They  can be derived from  the reduced dynamics, which under standard assumptions for thermalizing dynamics \cite{breuer_theory_2002} is described by a Markovian Master equation for the state  populations (also called Pauli rate equation): $\dot{\boldsymbol{p}}(t)=M\boldsymbol{p}(t)
	$.
	Here $\boldsymbol{p}$ is the energy level population vector, $M$ is the transition matrix with elements:
    \begin{equation}
        M_{ij}=\begin{cases}
       a_{ij}&i\neq j\\-\sum\limits_{k\neq j}a_{kj}&i=j 
    \end{cases}  
    \end{equation}
    $a_{kj}$ are the transition rates, and we assume that they keep detailed balance: $a_{jk}=a_{kj}e^{-\beta_{b}(E_j-E_k)}$, although this is not necessary for thermalization \cite{alicki2023violation}. Even though we focus on quantum systems, our results also apply to classical discrete Markovian systems. 
    
    The eigenvalues of $M$ are real ($0=\lambda_1\geq\lambda_2\geq...\geq\lambda_N$) and represent evolution speeds. Therefore, the system has up to $N-1$ distinct evolution speeds, which explain the Mpemba effect: a system farther away can evolve faster through ``larger distances''  if it has a fast speed component. We call  $\mathbf{V_i}$  the eigenvector corresponding to $\lambda_i$. The stationary state is thermal and corresponds to the eigenvector $\mathbf{V_1}=\Pi(\beta_b)\equiv Z^{-1}\{e^{-\beta_{b}E_1},...,e^{-\beta_{b}E_N}\}$. Here, $Z$ is the normalization factor.   Any initial state can be decomposed as a linear combination of the eigenvectors: 
	
	\begin{equation}      \mathbf{\boldsymbol{p}(0)}=\Pi(\beta_b)+\sum_{i=2}^N b_i \mathbf{V_i}.\label{eq:ini}
	\end{equation}
	\noindent	For $\lambda_2\neq0$, $|b_2|$ represents the overlap of the initial state with the slowest mode. In particular,  we will consider only initial thermal states, $\mathbf{\boldsymbol{p}(0)=\Pi(\beta_{in})}$.  As shown in \cite{lu2017nonequilibrium}, the Mpemba effect occurs iff $|b_2|$ has a maximum as a function of the initial inverse temperature, i.e., $\frac{d|b_2|}{d\beta_{in}}=0$ for $\beta_{in}\neq\beta_b$. For large systems, this calculation can be computationally demanding and may not yield a clear connection to the systems' physical parameters or underlying physical mechanisms.  Therefore, we seek alternative approaches to identify which types of systems exhibit the Mpemba effect.  To achieve this, we begin by studying the simplest system that can have more than one evolution speed and therefore may exhibit the Mpemba effect: a 3-level system.
	
	\section*{3-Level system necessary conditions}
	
	\begin{figure}
		\centering
		\includegraphics[width=1\linewidth]{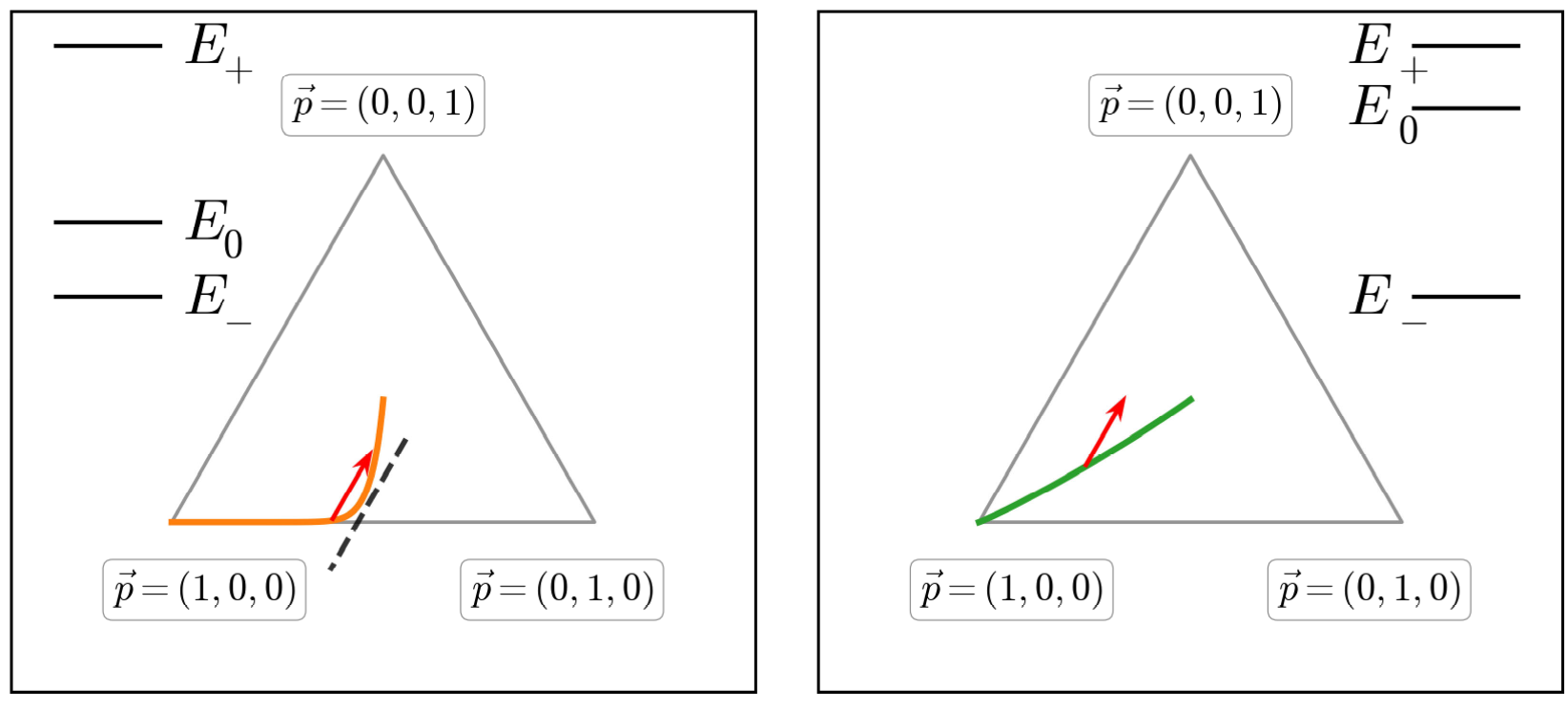}
		\caption{Testing the Mpemba effect for a 3LS formed from rotational energy-levels (left, $r>1$) vs a 3LS formed from hydrogen atom energy-levels (right, $r<1$). A system exhibits Mpemba if the fast eigenvector (red arrow) is parallel to a tangent of the quasistatic locus (orange/green line). As an illustration, we assume that the fast vector has the same angle in both cases. Here, the rotational case exhibits the Mpemba effect, and the hydrogen atom example does not. This is a consequence of the different energy-level spacing, quantified by $r$, which determines the shape of the quasistatic locus.}
		\label{fig:triangle}
	\end{figure}
	
 The advantage of studying a 3-level system is that it is simple enough to obtain analytical expressions for the eigenvectors and eigenvalues of  $\mathbf{M}$. Furthermore, it is possible to derive a geometric interpretation of the condition for the Mpemba effect: $\frac{d|b_2|}{d\beta_{in}}=0$. To explain this,  we note that any population state can be represented as a point in a 2D equilateral triangle. In this space, the quasistatic locus is a parameterized line that represents all the thermal states.  The Mpemba effect occurs iff the fast vector, $\mathbf{V_3}$, is parallel to a vector tangent to the quasistatic locus.  The exact shape of the quasistatic locus strongly depends on the relation among energy levels,  through the parameter $r\equiv\frac{\Delta_{+0}}{\Delta_{0-}}$ (see Figure \ref{fig:triangle}). Here $\Delta_{jk}=E_j-E_k$, where $E_+>E_0>E_-$ are the 3-level system energies. The angle between a tangent to the quasistatic locus and the triangle base ranges between $0$ and $arctan[\frac{1}{\sqrt{3}}(1+2r)]\leq \frac{\pi}{2}$. Therefore, a necessary condition for the Mpemba effect is a fast vector, $\pm \mathbf{V_3}$, with an angle between 0 and $\frac{\pi}{2}$ relative to the triangle base.  This imposes the following constraints on the transition rates depending on the value of $k\equiv\frac{a_{-+} +a_{0+}}{a_{-0}(1+e^{-\beta_b \Delta_{0-}})}$ (see the Supporting Information): 
	
	\begin{equation}
		\begin{aligned}
			a_{-+}C_{+-}-a_{0+}C_{+0} &> a_{-0} \left(1-e^{-\beta_b\Delta_{0-}}\right)&k\geq1 ; \\
			a_{-+}e^{-\beta_b\Delta_{0-}}&>a_{0+}&k\leq1 ,
            \label{eq:3lsc} 
		\end{aligned} 
	\end{equation}	
  	where $C_{ij}=1+\frac{1}{2}e^{-\beta_b\Delta_{ij}}$. $k$ determines whether the fastest dynamics is between the two lowest-energy states ($k<1$) or leaving the highest excited state ($k>1$). The fulfillment of any of the conditions above is necessary for the Mpemba effect and they become sufficient only for $r\rightarrow \infty$. The latter can be achieved in a system with nearly degenerate energy levels or by considering non-consecutive energy levels in an energy-increasing spectrum, such as rotational levels or a particle in a box. \eqref{eq:3lsc} shows that there are two different forms of producing the Mpemba effect:

    \begin{itemize}
        \item For $k>1$, the fast dynamic is leaving state $|+\rangle$. The whole condition can be fulfilled at any bath temperature if $a_{-+}\gg a_{-0},a_{0+}$.
        \item  For $k<1$, the fast dynamic is between states $|-\rangle$ and $|0\rangle$. Transition rates are always positive, so $a_{-+}e^{-\beta_b\Delta_{0-}}>a_{0+}$  can not be achieved at zero bath temperature. For non-zero temperatures, the whole condition  holds for $a_{-0}\gg a_{-+},a_{0+}$, and  the specific relation between $a_{-+}$ and $a_{0+}$ shown in \eqref{eq:3lsc}.
    \end{itemize}
    	
 Conditions (\ref{eq:3lsc}) may be underestimated because, in general, they are necessary but not sufficient.  Nevertheless, they have two main advantages relative to other conditions. Conditions (\ref{eq:3lsc})
 are simple enough to understand the mechanisms behind the Mpemba effect for 3-level systems and to determine when it does not occur. For example, they predict that a specific asymmetry in the rates is required for the Mpemba effect: $a_{-+}\sim a_{-0}\sim a_{0+}$ does not keep the conditions; 
	$a_{0+}\gg a_{-0},a_{-+}$  always precludes the Mpemba effect. Second, as we show below,  \eqref{eq:3lsc}  can also be used to study the Mpemba effect in non-degenerate N-level systems for which there are no simple expressions for the eigenvectors and eigenvalues of $\mathbf{M}$.
	
	\section*{Mpemba effect on an  $N-$level system}
 We develop a protocol for applying conditions (\ref{eq:3lsc}) to  $N-$level systems. The thermalization dynamics of  an $N-$level energy system is given by an $N\times N$ $\mathbf{M}$ matrix that includes  all the transition rates $a_{ij},$ $i,j \in \{1,...,N\}$.
	The problem is that conditions (\ref{eq:3lsc}) apply only to 3-level systems. To overcome this, we divide the $N-$level system energies in all the possible groups of 3-levels: $\{1,2,3\},\{1,2,4\},...\{1,2,N\},\{1,3,4\},...\{N-2,N-1,N\}$. We call each of these elements a triplet. To each triplet we associate the respective transition rates defined in $\mathbf{M}$, without modifying them (See the Supporting Information for a detailed example). This procedure differs from adiabatic elimination, in which, after a level is ignored, the transition rates are recalculated to get the effective ones.  Finally, for each triplet, we identified the highest-energy level as state $|+\rangle$, the intermediate level as state $|0\rangle$, and the lowest-energy level as state $|-\rangle$. At this point, we can apply conditions (\ref{eq:3lsc}) to each triplet and examine how they relate to the Mpemba effect in the $ N$- Level system. As shown below, if all triplets violate the conditions, the $ N$- level system will not exhibit the Mpemba effect. 
	We prove this analytically for a zero bath temperature and then study the non-zero-temperature case numerically.
	
	\subsection*{Analytical proof for zero-temperature baths}
  \, For zero-temperature baths, $\beta_b \rightarrow\infty$,  conditions (\ref{eq:3lsc}) simplify. $C_{ij}\rightarrow1$, so for $k>1$ we get $a_{-+}-a_{0+}>a_{-0}$. Since transition rates are positive, condition (\ref{eq:3lsc}) for $k<1$ is never satisfied.  Therefore, there are two ways of violating conditions (\ref{eq:3lsc}) : 
 
\begin{enumerate}[label=(\roman*)]
     \item $a_{-0}>a_{-+}+a_{0+}$ which implies $k<1$. 
     \item $a_{-+}+a_{0+}>a_{-0}>a_{-+}-a_{0+}$.
 \end{enumerate}
 
  Notice that (i) implies (ii) but not the other way around. That is why we call (i) a strong violation and (ii) a weak violation. Next, we analytically prove that strong violations exclude the Mpemba effect in an N-level system.
	
	\textit{Theorem:}
	Consider a non-degenerate $N-$level system with energies $E_1<...<E_N$ interacting with a thermal bath at zero temperature. Its dynamics is given by a master equation where  
	$\mathbf{M}$ is a triangular matrix, with a zero eigenvalue that we denote as $\Lambda_1$. The other eigenvalues are $\Lambda_i=\{-\sum_{k\neq i} a_{ki}\}$ for $i>1$.  For $i>1$, $i$ does not indicate the order of the eigenvalues' magnitude.  If $\mathbf{M}$ eigenvalues are not degenerate and for all the triplets, conditions (\ref{eq:3lsc}) are strongly violated (i.e., $a_{-0}>a_{-+}+a_{0+}$), then the $N-$level system does not present the Mpemba effect.

	\textit{Proof:} The transition matrix is a triangular matrix with non-degenerate eigenvalues. Then, the  ``non-ordered'' eigenvectors have the following structure:
	the element $k$ of the  eigenvector corresponding to $\Lambda_i$  is $v^{(i)}_k = 0$ for $k > i$ and $v^{(i)}_i \neq 0$.  As shown in the Supporting Information, if for all the triplets conditions (\ref{eq:3lsc}) are strongly violated, then the slowest non-zero eigenvalue is $\Lambda_N=-\sum_{k\neq N} a_{kN}$. Its corresponding eigenvector is the only one with a non-zero component for the $N$ coordinate. Therefore,  the expression for $b_2$ for a system initially in a thermal state with temperature $\beta_{in}$ simplifies to (see \eqref{eq:ini}):
	
	\begin{equation}
		b_2= \frac{e^{-\beta_{in}E_N}}{Z v^{(N)}_N},\label{eq:slow}
	\end{equation}
		where $Z=\sum_i e^{-\beta_{in}E_i}$. \eqref{eq:slow} has the same sign for all $\beta_{in}$ and its absolute value monotonically decreases with it (See the Supporting Information). Therefore, $\frac{d|b_2|}{d\beta_{in}}\neq0$, so the system does not experience the Mpemba effect.

	\subsection*{Numerics for non-zero temperature baths}
\begin{figure}
		\centering
		\includegraphics[width=1\linewidth]{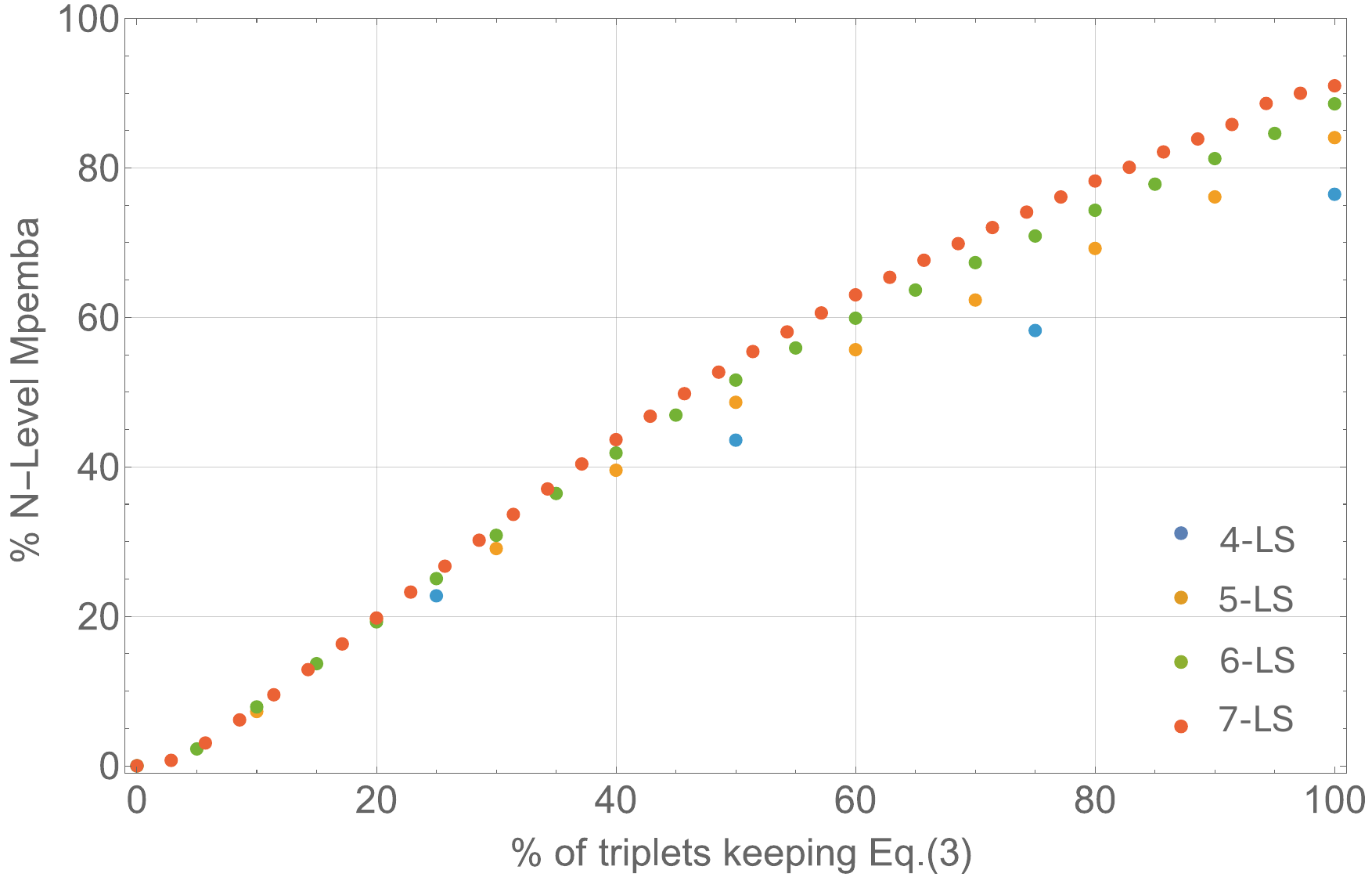}
		\caption{Probability for an $N-$level system to exhibit the Mpemba effect for random values of transition rates, energies and temperatures. The x-axis represents the percentage of triplets keeping  \eqref{eq:3lsc}. Convergence of the values was assessed by doubling the number of tested cases and requiring a change of less than $0.5\%$.  For the transition rates, we considered values from 0.05 to 1 and for $\beta_b E_i$ from $0$ to $1$.}
		\label{fig:nmpemba}
	\end{figure}

 \,Inspired by the theorem above, we use numerical simulations to test our protocol for baths at arbitrary temperatures and for baths at zero temperature for any violation of conditions (\ref{eq:3lsc}) (weak, strong or a combination of both). We randomly select positive values for $\beta_b$, the energy levels and the transition rates. We impose detailed balance on the latter  and order the energy levels as $E_1<...<E_N$. 
	As shown in Figure \ref{fig:nmpemba} for  4,5,6 and 7 level systems, if all the triplets violate conditions (\ref{eq:3lsc}), then the N-level system does not present the Mpemba effect.  Moreover, we have also tested fine-tuned cases that may be missed by random sampling, such as those with equal or zero transition rates. Also, in these cases, we do not find N-level Mpemba if none of the triplets satisfy conditions (\ref{eq:3lsc}). Based on this numerical result and our zero-temperature theorem, we \emph{conject} that \emph{if for an $N-$level system, all its triplets violate conditions (\ref{eq:3lsc}), then the $N-$level system does not experience the Mpemba effect.}  This conjecture implies that the Mpemba effect in any system can be studied by examining violations of the simple conditions for a 3-level system. As shown in the next section, this allows us to discard the Mpemba effect in several systems.  In addition to identifying the necessary physical mechanisms underlying the Mpemba effect, determining which systems \emph{do not} exhibit it is of great importance for machine learning algorithms.  In \cite{amorim2023predicting}, it was shown that neural networks trained exclusively on systems without the Mpemba effect could predict systems that experience the Mpemba effect. Our results allow us to generate examples that do not exhibit the Mpemba effect which can be used to train those algorithms.
	
Figure \ref{fig:nmpemba} shows other interesting trends.  The percentage of $N-$level Mpemba increases with the number of triplets keeping conditions (\ref{eq:3lsc}). It gets the highest value when all the triplets comply with \eqref{eq:3lsc}, although it does not reach 100\%. Moreover, for similar \% of triplets keeping \eqref{eq:3lsc}, the $N$-Level Mpemba percentage increases with the number of levels. Finally, the variance between the curves is lower for a low percentage of triplets complying with conditions (\ref{eq:3lsc}). In contrast to our conjecture, these results are limited to 7-level systems and may not reflect trends in larger systems.
	
	\section*{Discussion: Which systems present the Mpemba effect?}
	
 Once we have established the reliability of conditions (\ref{eq:3lsc}) for studying the Mpemba effect in an N-level system, we can use them to determine which systems exhibit the Mpemba effect and to investigate the underlying physical mechanisms. First, we focus on cases where \eqref{eq:3lsc} does not hold. As shown in the last section, this allows us to discard the $N-$level Mpemba.

 We start by considering different models for the functional dependence of the transition rate on the energy-level difference: $a_{jk}=f(\Delta)$, where $\Delta=E_j-E_k$. The exact form of $f$ depends on the details of the system in question. Still, it is a product of two components: a thermal component determined by the bath temperature and statistics (bosonic, fermionic, or classical), and the bath spectrum.  Notice that, as a consequence of the maximum entropy principle \cite{jaynes1957information}, the population of a thermal state follows an exponential distribution that decreases in energy. This makes the thermal component of $f$  a decreasing function of $\Delta$. As we explain below, this plays a central role in avoiding the Mpemba effect for several common bath spectra. 
 
 Because $C_{+0}>C_{+-}$, conditions (\ref{eq:3lsc}) are violated for $a_{-+}<a_{0+}$.
 The physical mechanism behind this violation can be better understood for the extreme case of  $a_{-+}\ll a_{0+}$ and zero bath temperature. 
 Under the above conditions, after a time of order  $t_0\sim 1/a_{0+}$ all the population of level $|+\rangle$ is transferred to $|0\rangle$, while the ground state population does not change.  At this point, only states $|0\rangle$ and $|-\rangle$ are populated and the rest of the dynamics is dominated by a \emph{single} timescale, $1/a_{-0}$. Therefore, the system that at $t_0$  is closer to equilibrium, will reach it first. For a zero bath temperature,  equilibrium corresponds to a state with all the population in the ground state. Hence, a system with less population in the ground state is farther away from equilibrium.   Because at $t_0$ the initially hotter system has less population in the ground state than the colder one,  it remains farther away from equilibrium and will not reach it first. This prevents the Mpemba effect. The analysis of \eqref{eq:3lsc} allows us to establish that this physical mechanism preventing the Mpemba effect is valid at any temperature and actually just requires $a_{-+}<a_{0+}$. As we show below, this simple mechanism may explain why the Mpemba effect is uncommon despite thermalization involving multiple evolution speeds.
 
The above physical mechanism implies that the Mpemba effect is prevented for  $\Delta$ where $f$ decreases. As an example of this behavior, we highlight the idealized bath with a constant spectrum. Due to the thermal component,  in this case $f(\Delta)$ is a decreasing function; therefore, a constant-spectrum bath does not support the Mpemba effect.  

Other mechanisms can be studied by considering extreme bath temperatures where \eqref{eq:3lsc} simplifies.  For high bath temperatures, $\beta_b\rightarrow 0$, the 3LS conditions are $a_{-+}>a_{0+}$, requiring only an increasing spectrum. For low bath temperatures, $\beta_b\rightarrow \infty$, the 3LS conditions are $a_{-+}>a_{0+}+a_{-0}$, therefore requiring a superadditive spectrum and automatically discarding any concave $f$ for Mpemba for low  temperature baths.

To study the Mpemba effect at any bath temperature and for structured spectra, we focus on specific bath spectra.  Typical examples include:  sub-Ohmic, Ohmic and super-Ohmic. A phenomenological cutoff should bound these spectra to prevent unphysical divergences \cite{weiss2012quantum}.   The presence of the cutoff as well as the bath statistic force $f$ to be a monotonically decreasing function for high $\Delta$ and therefore forbids the Mpemba at high frequencies. 
	
	 In particular, we choose a bosonic bath with  the following functional form for the transition rates \cite{weiss2012quantum}:

	\begin{equation}
		a_{kj}=f(\Delta)\propto \Delta^{s-1}e^{-\Delta/\Delta_c}\left(\frac{1}{e^{\beta_b \Delta}-1}+1\right) \label{eq:model}
	\end{equation}  
		where  $\Delta_c$ is a cutoff frequency. Here, we use an exponential cutoff, but we did not observe qualitative differences when we tested our results with other standard functional cutoffs, such as the Drude regularization. $s$ depends on the physical realization and determines the spectrum type \cite{weiss2012quantum}. For  $s<1$ the spectrum is called sub-Ohmic, $s=1$  Ohmic and $s>1$ super-Ohmic.  The need for transition rates that increase with $\Delta$ automatically discards sub-Ohmic and Ohmic spectra for $N$-level Mpemba. 
       These spectra describe multiple thermalization scenarios, and the fact that they do not produce the Mpemba effect potentially explains its rarity.  Moreover, for $s\leq2$ it can be analytically shown that the $N$-Level Mpemba effect does not occur at high and low bath temperatures (See the Supporting Information), and we checked numerically that this result holds for any temperature. Therefore, the only transition rates  of the form \eqref{eq:model} that can produce Mpemba are 
       those based on  super-Ohmic spectra with $s>2$.  A distinctive physical behavior for super-Ohmic spectra with $s$ below and above $2$ has also been found in other cases, such as dissipative two-level systems \cite{leggett1987dynamics},  long time survival of coherence \cite{giraldi2013survival} and other thermodynamic anomalies \cite{adamietz2014thermodynamic}.
	
\begin{figure}[htbp]
		\centering
		\includegraphics[width=1\linewidth]{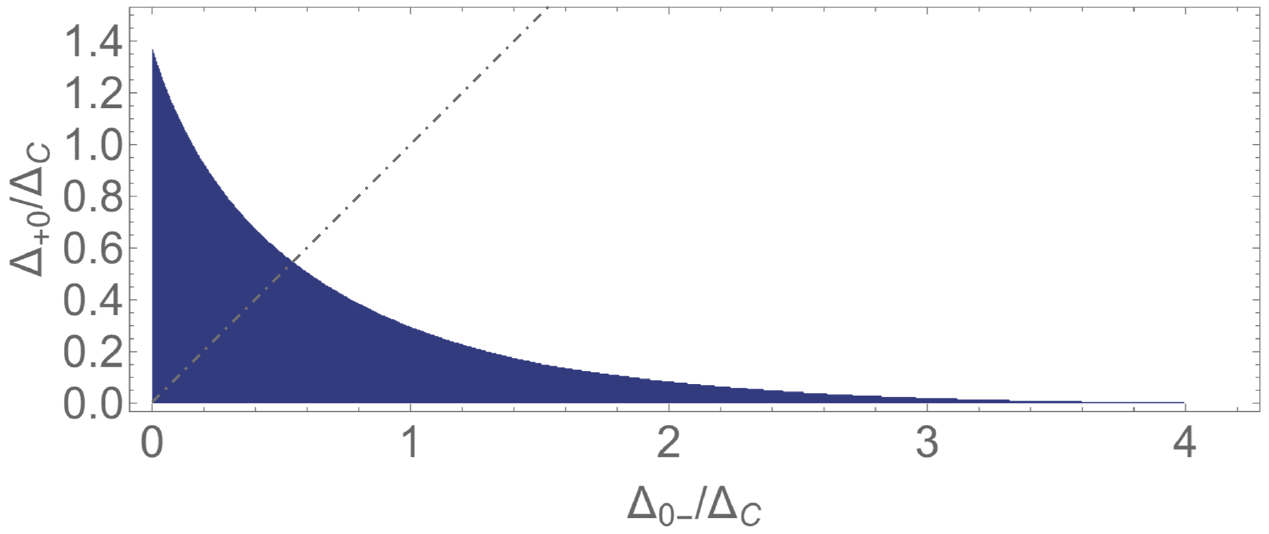}
		\caption{Testing the Mpemba necessary conditions for a 3LS as function of the energy level spacing $\Delta_{jk}=E_j-E_k$. The blue region indicates compliance with conditions (\ref{eq:3lsc}). Therefore, $N-$level systems with triplets exclusively in the white regions will not exhibit the Mpemba effect.  Above (below) the diagonal dashed line, the energy levels correspond to increasing (decreasing) energy-level spacing.  The plot shows an asymmetry favoring systems with decreasing energy-level spacing ($r<1$, e.g., hydrogen energy levels) for the conditions (\ref{eq:3lsc}). Plot parameters: super-Ohmic bath spectra with $s=3$, $\beta_b \Delta_c=1$.}
		\label{fig:s3}
	\end{figure}

	Having $s>2$ is not sufficient to ensure compliance with the 3LS conditions (\ref{eq:3lsc}). As shown by the blue regions in Figure \ref{fig:s3}, the conditions are met only up to a certain set of differences among the system energy levels. Therefore, the Mpemba effect can be ruled out for larger energy differences. One can estimate the order of magnitude
	for the energy difference to keep conditions (\ref{eq:3lsc}) to coincide with the maximum of $f$ for $\beta_b\rightarrow0$, that is $|\Delta_{kj}|\lesssim(s-2)\Delta_C$.

\section*{Conclusions}
\eqref{eq:3lsc} represents simple conditions for the Mpemba effect in a Markovian 3LS. Furthermore,  we prove analytically for baths at zero temperature, and show numerically for baths at any temperature, that the 3LS conditions can be used to study the effect in an $N$-level system. These conditions require transition rates that increase with the energy difference. Due to the maximum entropy principle, the thermal component of the transition rates decays with the energy difference, limiting the bath spectra that can produce the Mpemba effect. This allows us to rule out the $N$-level system Mpemba effect for sub-Ohmic, Ohmic, and super-Ohmic spectra with $s\leq2$ and limit it to specific energy-level separations for spectra with $s>2$. By excluding all these cases, we helped explain why the Mpemba effect remains a thermodynamic anomaly despite the presence of multiple time scales during thermalization. Moreover, our results allowed us to investigate the physical mechanisms underlying the Mpemba effect and to provide a clear path to determine which types of systems exhibit it. Finally, \eqref{eq:3lsc}  could help identify relevant parameters for Mpemba-related experiments and numerical simulations, as well as for diverse applications such as heating/cooling protocols, sensors, quantum state preparation, and Monte Carlo algorithms.\\
 
\textit{Acknowledgement.}
We thank Sergiy Denysov, Dariusz Chruscinski, Eviatar Procaccia and Saar Rahav for useful discussions. D.G.K.~is  supported by the ISRAEL SCIENCE FOUNDATION (grant No. 2247/22) and by the Council for Higher Education Support Program for Hiring Outstanding Faculty Members in Quantum Science and Technology in Research Universities.

\onecolumngrid

\setcounter{equation}{0}
\renewcommand{\theequation}{S\arabic{equation}}

\setcounter{figure}{0}
\renewcommand{\thefigure}{S\arabic{figure}}

\setcounter{section}{0}
\renewcommand{\thesection}{S\arabic{section}}

\setcounter{subsection}{0}
\renewcommand{\thesubsection}{\thesection.\arabic{subsection}}
\vspace{1cm}

\begin{centering}
{\large \bf Supporting Information}\\
\end{centering}

\section*{Necessary conditions for the Mpemba effect in a 3-level system}

\begin{figure}[H]
    \centering
    \includegraphics[width=0.5\linewidth]{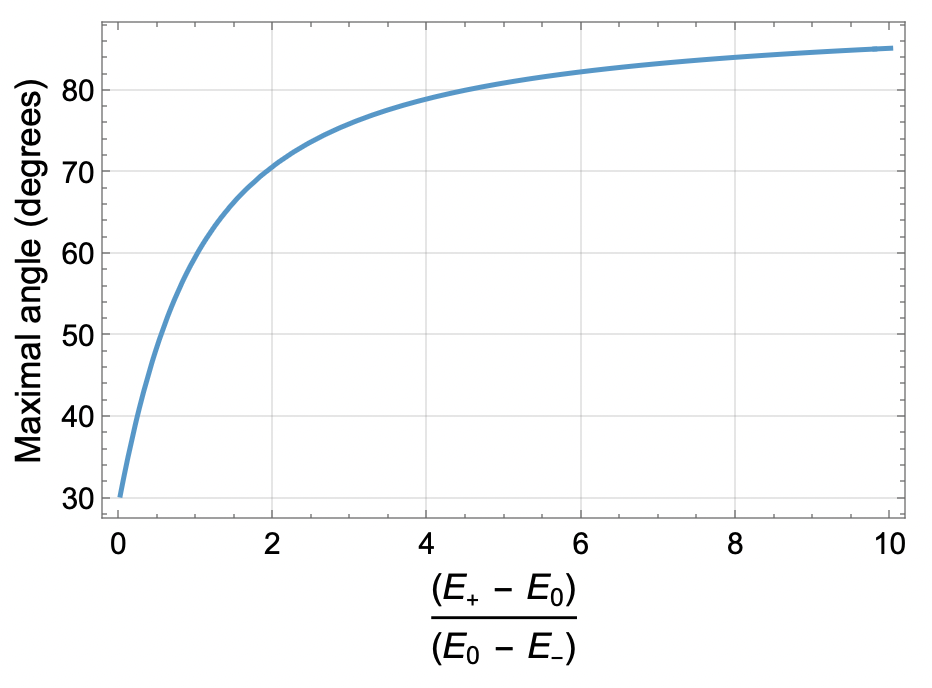}
    \caption{Maximum angle between the tangent vector and the base of the triangle as a function of the energy gap ratio \eqref{e relation}.}
    \label{relation graph}
\end{figure}

In \cite{lu2017nonequilibrium}, it was shown that the Mpemba effect happens when the fast vector is parallel to a vector that is tangent to the quasistatic locus (See Fig. 1 in the main text).
To explore this, we map the 3-level system (3LS) state space onto a 2D triangle, where each vertex corresponds to a state with a single energy level populated.  The tangent to the quasistatic locus has an angle (relative to the triangle base) that ranges from zero to a maximum that depends on the ratio of energy-level spacings. Specifically, for energy levels $E_-<E_0<E_+$, the ratio is defined as: 
\begin{align}\label{e relation}
    r=\frac{\Delta_{+0}}{\Delta_{0-}}=\frac{E_+-E_0}{E_0-E_-},
\end{align}
where $\Delta_{ij}=E_i-E_j$. The maximal angle of the tangent vector can not pass  $\pi/2$ (see Fig. \ref{relation graph}). Therefore, a necessary condition for the Mpemba effect is that the fast vector forms an angle between 0 and $\pi/2$ with relation to the base of the triangle. An angle between $\pi$ and $3\pi/2$ will also produce the Mpemba effect. Below, we present how the condition in the fast vector angle translates into conditions on the transition rates.\\

\noindent Consider the 3LS transition matrix:
\begin{align} 
    M_{3LS}=
    \begin{pmatrix}
        -a_{0-}-a_{+-}&a_{-0}&a_{-+}\\
        a_{0-}&-a_{-0}-a_{+0}&a_{0+}\\
        a_{+-}&a_{+0}&-a_{-+} - a_{0+}\\
    \end{pmatrix}.
\end{align}
Moreover, we assume the transition rates keep detailed balance $\frac{a_{ij}}{a_{ji}}=e^{-\beta\Delta_{ij}}$.\\
To map population states and vectors to the 2D triangular plane, we use the following rotation matrix: 
\begin{align}
 R=   \begin{pmatrix}
    -\dfrac{1}{\sqrt{2}} & \dfrac{1}{\sqrt{2}} & 0 \\
    -\dfrac{1}{\sqrt{6}} 
    & -\dfrac{\sqrt{\frac{3}{2}}}{2} + \dfrac{1}{2\sqrt{6}} 
    & \dfrac{\sqrt{\frac{3}{2}}}{2} + \dfrac{1}{2\sqrt{6}} \\
    \dfrac{1}{\sqrt{3}} & \dfrac{1}{\sqrt{3}} & \dfrac{1}{\sqrt{3}}
    \end{pmatrix}.
\end{align}
The rotated fast vector is found to  be 
\begin{align}
   R  \bf{V_3}=
    \begin{pmatrix}
        \frac{-a_{-0}+a_{-+}+a_{0+}-a_{0-}-\sqrt{(-a_{-0}+a_{-+}+a_{0+}-a_{0-})^2-(a_{+0}-a_{+-})(-2(a_{-0}-a_{-+}+a_{0+})+2a_{0-}+a_{+-}-a_{+0})}}{\sqrt{2}(a_{+0}-a_{+-})},\sqrt{\frac{3}{2}},0
    \end{pmatrix}.
\end{align}
For the vector to form an angle between 0 and $\pi/2$ w.r.t the base of the triangle, the first component of the vector must be larger than 0.\\
We consider two cases:

\noindent \textbf{Case 1}

\noindent The first is for $a_{+0}>a_{+-}$.
For the term outside the root square to be positive, we need:\\
\begin{align}
    a_{-+}+a_{0+}>a_{-0}+a_{0-}.
\end{align}

\noindent Moreover, if
\begin{align}\label{cond}
    (-2(a_{-0}-a_{-+}+a_{0+})+2a_{0-}+a_{+-}-a_{+0})>0,
\end{align}
 the root will be smaller than the expression outside and the vector component will be positive. Rewriting the above inequality we get: 

\begin{align}\label{final cond}
   a_{-+}+\frac{1}{2}a_{+-}>a_{-0}+a_{0+}-a_{0-}+\frac{1}{2}a_{+0}.
\end{align}

\noindent \textbf{Case 2}

\noindent The second case is for $a_{+0}<a_{+-}$. 
The first component of the vector can be rewritten as
{\scriptsize \begin{align}
    \frac{-(-a_{-0}+a_{-+}+a_{0+}-a_{0-})+\sqrt{(-a_{-0}+a_{-+}+a_{0+}-a_{0-})^2+(a_{+-}-a_{+0})(-2(a_{-0}-a_{-+}+a_{0+})+2a_{0-}+a_{+-}-a_{+0})}}{\sqrt{2}(a_{+-}-a_{+0})}.
\end{align}}
If ${-a_{-0}+a_{-+}+a_{0+}-a_{0-}<0}$ the vector component will always be positive and the angle will always be under $\pi/2$, and for $-a_{-0}+a_{-+}+a_{0+}-a_{0-}>0$ we need again to satisfy condition \eqref{cond} that leads to condition \eqref{final cond}\\

\noindent For $-a_{-0}+a_{-+}+a_{0+}-a_{0-}=0$, \eqref{final cond} coincides with $a_{+0}<a_{+-}$.

\noindent Next, we define the parameter
\begin{align*}
    k\equiv \frac{a_{-+}+a_{0+}}{a_{-0}+a_{0-}} = \frac{a_{-+}+a_{0+}}{a_{-0}(1+e^{-\beta_b \Delta_
    {0-}})}, 
\end{align*}
where we used detailed balance in the second equality.  Using detailed balance, we rewrite the necessary conditions for the 3LS Mpemba effect as:
\begin{equation}
		\begin{aligned}
			a_{-+}C_{+-}-a_{0+}C_{+0} &> a_{-0} \left(1-e^{-\beta_b\Delta_{0-}}\right)&k\geq1;  \\
			a_{-+}e^{-\beta_b\Delta_{0-}}&>a_{0+}&k\leq1, \label{eq:3lscss} 
		\end{aligned} 
	\end{equation}
where $\Delta_{ij} =  \epsilon_i-\epsilon_j$ and $C_{ij}=1+\frac{1}{2}e^{-\beta_b\Delta_{ij}}$.

\section*{$N-$level system protocol}
In this section, we provide a detailed example of how our protocol for testing the Mpemba effect in an $ N$-level system works. The example below is for a 4-level system. Its generalization to an $N$-Level system is straightforward.

\noindent Consider  a 4LS with ordered energy levels $E_{1}<...<E_{4}$. Its transition matrix is 

\begin{equation}
M=\left(\begin{array}{cccc}
-a_{41}-a_{31}-a_{21} & a_{12} & a_{13} & a_{14}\\
a_{21} & -a_{12}-a_{32}-a_{42} & a_{23} & a_{24}\\
a_{31} & a_{32} & -a_{13}-a_{23}-a_{43} & a_{34}\\
a_{41} & a_{42} &a_{43} & -a_{14}-a_{24}-a_{34}
\end{array}\right).\label{eq:m4}
\end{equation}
Furthermore, we assume that the transition rates keep detailed balance. We now find all the triplets. That is, all possible groups of size 3 formed from the energy levels of the 4LS. We get the following
groups: A) \{1, 2, 3\}, B) \{1, 2, 4\}, C) \{1, 3, 4\} and D) \{2, 3, 4\}.

\subsection*{3LS conditions for group A}
For group A, $E_+=E_3$, $E_0=E_2$ and $E_-=E_1$. Therefore,

\begin{align*}
    k = \frac{a_{13}+a_{23}}{a_{12}(1+e^{-\beta_b \Delta_
    {21}})} 
\end{align*}
and the 3LS conditions are

\begin{equation}
		\begin{aligned}
			a_{13}C_{31}-a_{23}C_{32} &> a_{12} \left(1-e^{-\beta_b\Delta_{21}}\right)&k\geq1;  \\
			a_{13}e^{-\beta_b\Delta_{21}}&>a_{23}&k\leq1, 
            \label{eq:3lsca} 
		\end{aligned} 
	\end{equation}	
  	where $C_{ij}=1+\frac{1}{2}e^{-\beta_b\Delta_{ij}}$. The transition rates are the same as in \eqref{eq:m4}.

\subsection*{3LS conditions for group B}
For group B, $E_+=E_4$, $E_0=E_2$ and $E_-=E_1$. Therefore,

\begin{align*}
    k = \frac{a_{14}+a_{24}}{a_{12}(1+e^{-\beta_b \Delta_
    {21}})} 
\end{align*}
and the 3LS conditions are

\begin{equation}
		\begin{aligned}
			a_{14}C_{41}-a_{24}C_{42} &> a_{12} \left(1-e^{-\beta_b\Delta_{21}}\right)&k\geq1;  \\
			a_{14}e^{-\beta_b\Delta_{21}}&>a_{24}&k\leq1,
            \label{eq:3lscb} 
		\end{aligned} 
	\end{equation}	
  	where $C_{ij}=1+\frac{1}{2}e^{-\beta_b\Delta_{ij}}$. The transition rates are the same as in \eqref{eq:m4}.

\subsection*{3LS conditions for group C}
For group C, $E_+=E_4$, $E_0=E_3$ and $E_-=E_1$. Therefore,

\begin{align*}
    k = \frac{a_{14}+a_{34}}{a_{13}(1+e^{-\beta_b \Delta_
    {31}})} 
\end{align*}
and the 3LS conditions are

\begin{equation}
		\begin{aligned}
			a_{14}C_{41}-a_{34}C_{43} &> a_{13} \left(1-e^{-\beta_b\Delta_{31}}\right)&k\geq1;  \\
			a_{14}e^{-\beta_b\Delta_{31}}&>a_{34}&k\leq1, 
            \label{eq:3lscc} 
		\end{aligned} 
	\end{equation}	
  	where $C_{ij}=1+\frac{1}{2}e^{-\beta_b\Delta_{ij}}$. The transition rates are the same as in \eqref{eq:m4}.

\subsection*{3LS conditions for group D}
For group D, $E_+=E_4$, $E_0=E_3$ and $E_-=E_2$. Therefore,

\begin{align*}
    k = \frac{a_{24}+a_{34}}{a_{23}(1+e^{-\beta_b \Delta_
    {32}})} 
\end{align*}
and the 3LS conditions are

\begin{equation}
		\begin{aligned}
			a_{24}C_{42}-a_{34}C_{43} &> a_{23} \left(1-e^{-\beta_b\Delta_{32}}\right)&k\geq1;  \\
			a_{24}e^{-\beta_b\Delta_{32}}&>a_{34}&k\leq1, 
            \label{eq:3lscd} 
		\end{aligned} 
	\end{equation}	
  	where $C_{ij}=1+\frac{1}{2}e^{-\beta_b\Delta_{ij}}$. The transition rates are the same as in \eqref{eq:m4}.

\section*{$N-$level system: analytical proof for zero-temperature baths}

In this section, we prove that at zero temperature, one can discard the Mpemba effect in an $ N$-level system if the 3LS conditions are strongly violated for all $ N$-level triplets. We prove it by induction. First, we show that the protocol works for a 4-level system (4LS), and then we show that, by assuming it works for an $N$-Level system, it also works for an $N+1$-Level system.

\subsection*{4LS}

Consider the transition matrix of a 4LS interacting with a bath at zero temperature. We assume that the energy levels are ordered: $E_{1}<...<E_{4}:$

\begin{equation}
M=\left(\begin{array}{cccc}
0 & a_{12} & a_{13} & a_{14}\\
0 & -a_{12} & a_{23} & a_{24}\\
0 & 0 & -a_{13}-a_{23} & a_{34}\\
0 & 0 & 0 & -a_{14}-a_{24}-a_{34}
\end{array}\right).
\end{equation}
Because this is a triangular matrix, its eigenvalues are just the diagonal elements.
elements: $\{0,-a_{12},-a_{13}-a_{23},-a_{14}-a_{24}-a_{34}\}.$ Notice that the last three eigenvalues are not ordered by value. Call
$\{v^{(1)},v^{(3)},v^{(3)},v^{(4)}\},$ their  respective eigenvectors.
Furthermore, we assume that the eigenvalues are non-degenerate. Then,
the eigenvectors have the following structure: $v_{k}^{(i)}=0$ for
$k>i$ and $v_{i}^{(i)}\neq0$, where $v_{k}^{(i)}$ is the k coordinate
of vector $i$$.$ The initial conditions of a thermal state at inverse
temperature $\beta_{in}$ are found by solving the following equation

\begin{equation}
\boldsymbol{p}(0)=\Pi(\beta_{b}\rightarrow\infty)+\sum_{i=2}^{4}b_{i}^{not-ord}v^{(i)},
\end{equation}
where for zero temperature $\Pi(\beta_{b}\rightarrow\infty)=\{1,0,0,0\}$
and $p(0)=\{e^{-\beta_{ini}E_{1}},e^{-\beta_{ini}E_{2}},e^{-\beta_{ini}E_{3}},e^{-\beta_{ini}E_{4}}\}/Z$.
$Z$ is a normalization constant. Here $b_{i}^{not-ord}$ represents
the overlap with the eigenvector $v^{(i)}$. Here, the slowest mode is not necessarily  $i=2$. It
could be any of the eigenvalues as long as $i>1$. 

\noindent Notice that $v^{(4)}$ is the only eigenvector with a non-zero 4th
component. If the slowest vector is $v^{(4)}$, then the overlap of
the initial state with the slow degree of motion is

\begin{equation}
b_{4}^{not-ord}=\frac{e^{-\beta_{in}E_{4}}}{Zv_{4}^{(4)}}.
\end{equation}

\noindent The sign of $b_{4}^{not-ord}$ is the same as the sign of $v_{4}^{(4)}$
and does not depend on $\beta_{in}$. Its derivative is:

\begin{gather*}
\frac{d|b_{4}^{not-ord}|}{d\beta_{in}}=\frac{1}{|v_{4}^{(4)}|}\frac{d}{d\beta_{in}}\frac{e^{-\beta_{in}E_{4}}}{Z}\propto\frac{1}{|v_{4}^{(4)}|}\left(-E_{4}e^{-\beta_{in}E_{4}}\sum_{k=1}^{4}e^{-\beta_{in}E_{k}}+e^{-\beta_{in}E_{4}}\sum_{k=1}^{4}E_{k}e^{-\beta_{in}E_{k}}\right)=\\
\frac{1}{|v_{4}^{(4)}|}e^{-\beta_{in}E_{4}}\sum_{k=1}^{4}(E_{k}-E_{4})e^{-\beta_{in}E_{k}}<0.
\end{gather*}

\noindent This is always negative.  This implies that
if $v^{(4)}$ is the slowest eigenvector, then the 4LS does not experience the Mpemba effect. To prove that $v^{(4)}$ is the slowest eigenvector, we need to show two things:

\begin{equation}
a_{14}+a_{24}+a_{34}<a_{12};\label{eq:4lsa}
\end{equation}

\begin{equation}
a_{14}+a_{24}+a_{34}<a_{13}+a_{23}.\label{eq:4lsb}
\end{equation}

\noindent To show this, we first need to understand how the 3LS conditions can be violated. Consider a non-degenerate 3LS with energies $\{E_{+}>E_{0}>E_{-}\}.$
At zero temperature, two conditions imply no Mpemba effect
in the 3LS (each of them is a sufficient condition  by itself for no Mpemba):

a) Strong violation 
\begin{equation}
a_{-0}>a_{-+}+a_{0+}.\label{eq:A1-1}
\end{equation}

b) Weak violation

\begin{equation}
a_{-0}>a_{-+}-a_{0+}.\label{eq:A1-1-1}
\end{equation}
Notice that if the strong violation is satisfied, automatically the weak one is satisfied, but the other way around does not hold. 

\noindent We now find all the triplets. That is, all the possible groups formed by 3 energy levels of the
4LS. We get the following groups: A) $\{1,2,3\}$, B) $\{1,2,4\},$C)
$\{1,3,4\}$ and D) $\{2,3,4\}$. 

\noindent The strong violation of group A is 

\[
a_{12}>a_{13}+a_{23}.
\]
Using this inequality if we prove \ref{eq:4lsb}, then \ref{eq:4lsa}
automatically holds. 

\noindent The strong violations of groups C and D are 

\begin{equation}
a_{13}>a_{14}+a_{34};\label{eq:C1-1}
\end{equation}

\begin{equation}
a_{23}>a_{24}+a_{34}.\label{eq:D1-1}
\end{equation}
Adding these conditions, we get 

\[
a_{13}+a_{23}>a_{14}+a_{24}+2a_{34}>a_{14}+a_{24}+a_{34},
\]
where in the second inequality we use that the transition rates are
positive. This proof \ref{eq:4lsb} and therefore shows that $v^{(4)}$
is the slowest eigenvector. Because $|b_{4}^{not-ord}|$ does not
have a maximum as a function of $\beta_{in}$, the system does not present
the Mpemba effect.

\subsection*{$N+1-$ level system}

We assume that for $N$ our protocol holds, and check if it holds for
$N+1$. We consider $\beta_{b}\rightarrow\infty$. The transition
matrix is triangular and has the same properties as in the 4LS case.
The eigenvalues of the matrix are $\{0,-a_{12},-a_{13}-a_{23},....,-\sum_{i=1}^{N}a_{iN+1}\}$. Furthermore,
assuming no degeneracy of these eigenvalues, the eigenvectors $v^{(i)}$ have zeros for $k>i$
and $v_{i}^{(i)}\neq0$.

\noindent The initial conditions are found by solving the following equation:

\begin{equation}
\mathbf{p}(0)=\Pi(\beta_{b}\rightarrow\infty)+\sum_{i=2}^{N+1}b_{i}^{not-ord}v^{(i)},
\end{equation}
where for zero temperature $\Pi(\beta_{b}\rightarrow\infty)=\{1,0,...,0\}$,
$p(0)=\{e^{-\beta_{in}E_{1}},...,e^{-\beta_{in}E_{N+1}}\}/Z$ . $Z$
is a normalization constant and $E_{1}<...<E_{N+1}$. If the slowest
vector is $v^{(N+1)},$ then

\begin{equation}
b_{N+1}^{not-ord}=\frac{e^{-\beta_{in}E_{N+1}}}{Zv_{N+1}^{(N+1)}}.
\end{equation}

Following similar arguments as for the 4LS,

\begin{gather*}
\frac{d|b_{N+1}^{not-ord}|}{d\beta_{in}}=\frac{1}{|v_{N+1}^{(N+1)}|}\frac{d}{d\beta_{in}}\frac{e^{-\beta_{in}E_{4}}}{Z}\propto\frac{1}{v_{N+1}^{(N+1)}|}\left(-E_{N+1}e^{-\beta_{in}N+1}\sum_{k=1}^{N+1}e^{-\beta_{in}E_{k}}+e^{-\beta_{in}E_{N+1}}\sum_{k=1}^{N+1}E_{k}e^{-\beta_{in}E_{k}}\right)=\\
\frac{1}{|v_{N+1}^{(N+1)}|}e^{-\beta_{in}E_{N+1}}\sum_{k=1}^{N+1}(E_{k}-E_{N+1})e^{-\beta_{in}E_{k}}<0,
\end{gather*}
which is always negative. Therefore, it has no maximum, so it does not exhibit the Mpemba effect. So if the slowest
eigenvector is $v^{(N+1)}$ there is no Mpemba. 

The fact that we are assuming that our protocol holds for $N$ implies
that if the strong violation holds for all the triplets
formed from $N$ energy levels, then the slowest eigenvalue for the
$N$ case is $-\sum_{i=1}^{N-1}a_{iN}$. So we just need to show that
a strong violation of the ``new'' 3LS conditions (those involving
level $N+1)$ will imply that $\sum_{i=1}^{N-1}a_{iN}>\sum_{i=1}^{N}a_{iN+1}$.
We focus on the 3-level groups of the form $\{i,N,N+1\}$. For this,
the strong violation implies

\begin{equation}
a_{iN}>a_{iN+1}+a_{NN+1}.
\end{equation}
Summing over $i$ (and remembering that $N>2)$

\begin{equation}
\sum_{i=1}^{N-1}a_{iN}>\sum_{i=1}^{N-1}a_{iN+1}+(N-1)a_{NN+1}>\sum_{i=1}^{N}a_{iN+1}.
\end{equation}
This implies that if the strong violation holds for the triplets
$\{i,N,N+1\}$, and assuming that our protocol holds for N, $v^{(N+1)}$
will be the slowest eigenvector. Therefore, the $ N+1$-Level system does not exhibit the Mpemba effect.

\section*{Transition rates for  $s=2$ }
In this section, we prove that at high/low temperatures the transition rates for $s=2$ do not satisfy the 3LS conditions.  We assume that the transition rates have the form

	\begin{equation}
		a_{kj}=f(\Delta)\propto \Delta^{s-1}e^{-\Delta/\Delta_c}(1/(e^{\beta_b \Delta}-1)+1) .\label{eq:models}
	\end{equation}

    \subsection*{High temperature}
For $\beta_b\rightarrow 0$, the transition rates for $s=2$ are

\begin{equation}
a_{kj}\propto\frac{1}{\beta_b}e^{-\Delta/\Delta_c}.    
\end{equation} 
This is a decreasing function of $\Delta$. As discussed in the main text,   decreasing functions do not produce the Mpemba effect.

\subsection*{Low temperatures}
For $\beta_b\rightarrow \infty$ we prove by contradiction that the transition rates do not keep the 3LS condition. The transition rates for $s=2$ are

\begin{equation}
a_{kj}\propto \Delta e^{-\Delta/\Delta_c},
\label{eq:lowt}
\end{equation} 
and the condition for Mpemba is $a_{-+}>a_{0+}+a_{-0}$.   Substituting in this inequality, keeping the 3LS conditions implies 

 \begin{equation}
  \Delta_{+-} e^{-\Delta_{+-}/\Delta_c}>\Delta_{+0} e^{-\Delta_{+0}/\Delta_c}+\Delta_{0-} e^{-\Delta_{0-}/\Delta_c}.
   \end{equation}
   Here, $\Delta_{ij}=E_i-E_j$. Rewriting the above inequality we get
\begin{equation}
     \Delta_{+-} >\Delta_{+0} e^{\Delta_{0-}/\Delta_c}+\Delta_{0-} e^{\Delta_{+0}/\Delta_c}.
   \end{equation}
   Next, we use that the exponentials on the right-hand side are larger than 1 and all the $\Delta_{ij}>0$, therefore,
\begin{equation}
     \Delta_{+-} >\Delta_{+0}+\Delta_{0-} =\Delta_{+-}.
   \end{equation}
This is a contradiction. Therefore, our initial assumption that the transition rates keep the 3LS conditions is incorrect.

\end{document}